# Oxide-apertured microcavity single-photon emitting diode

D. J. P. Ellis, A. J. Bennett, and A. J. Shields,

*Toshiba Research Europe Limited, Cambridge Research Laboratory, 260 Science Park, Milton Road, Cambridge, CB4 OWE, UK.*

P. Atkinson, and D. A. Ritchie

*Cavendish Laboratory, Cambridge University, JJ Thomson Avenue, Cambridge, CB3 0HE, UK.*

**Abstract**

We have developed a microcavity single-photon source based on a single quantum dot within a planar cavity in which wet-oxidation of a high-aluminium content layer provides lateral confinement of both the photonic mode and the injection current. Lateral confinement of the optical mode in optically pumped structures produces a strong enhancement of the radiative decay rate. Using microcavity structures with doped contact layers, we demonstrate a single-photon emitting diode where current may be injected into a single dot.

In recent decades, advances in the manufacture of microstructures have provided the means to control and trap light within wavelength-scale structures. An example is the vertical-cavity surface-emitting laser (VCSEL) which has been widely studied[1] as a promising route to high-efficiency, fast and low-threshold devices for tele- and data-communications. The use of quantum dot layers and improved cavity designs has lead to improved VCSEL performance[2]. More recently a generation of light sources based on the quantum-mechanically discrete states of single quantum dots have appeared which are able to generate photon streams with non-classical statistics[3], or entangled photon pairs[4]. Attempts to increase the efficiency of these sources has thus far been focused on pillar microcavities[5,6] and defect photonic crystals[7]. In these cases the optical mode is strongly confined laterally, either by the large refractive index contrast between the semiconductor and surrounding air, or by the photonic crystal. However, the rough exposed surfaces of these structures scatter photons, reducing the cavity mode quality factors, Q and the proportion of emission that may be collected[8].

Here we detail a single photon source with a cavity in which the optical mode is spatially separated from any etched surface[9]. A cavity confines light vertically between two Bragg mirrors and laterally by an aluminium oxide layer produced by the wet oxidation of aluminium-rich $Al_xGa_{1-x}As$[8,10]. The emission from the fundamental mode can be collected easily and the design can be simply adapted for electrical injection. We observe that the light confinement can be strong enough to enhance the single photon collection efficiency via the Purcell effect. We then demonstrate electroluminescence from a single quantum dot in an electrically driven structure. A schematic of the device design used is shown in figure 1.

The degree of optical confinement afforded by such oxide-confined structures is dependent upon a number of parameters, including the vertical position, thickness and cross-sectional profile of the oxide layer. These parameters all change the effective refractive index distribution in the structure and have been studied in the context of VCSEL design[11,12]. Here, we employ a $\lambda/2$ thick oxide layer with a tapered profile to give a 0.18 change in effective refractive index between the fully oxidised ($n_{eff} = 3.09$) and unoxidised ($n_{eff} = 3.27$) sections of the wafer. The inner edge of the oxide annulus was placed at an electric field node. The calculated radial index variation in this structure is shown in the lower part of figure 1. A tapered oxide profile was used to reduce photon scattering from the oxide region[11]. The AlOx annulus not only provides lateral confinement to the mode but can also act as a current aperture, allowing the active area of an electrical device to be reduced[13].

Wafer 1 consisted of 17 (25) period GaAs/AlGaAs distributed Bragg reflectors (DBRs) above (below) a $\lambda$-thick GaAs cavity, which contained a low density of self-assembled InAs quantum dots at its center. The oxidation region, of the type described by Hegblom *et al*,[14] was placed between the cavity and the upper mirror. Mesas were etched and subsequently oxidised for ~ six minutes at 400°C to produce a ring of oxide penetrating in 5 - 6 $\mu$m from the edge of each mesa.

Figure 2(a) shows a typical photoluminescence (PL) spectrum recorded from a 12 $\mu$m diameter mesa with an excitation laser power of ~ 0.1 mW at ~ 4 K. A clear mode structure was observed in all pillars of this type, which was blueshifted relative to the planar cavity mode. The origin and assignment of these modes is well known from work on optical fibres with circular cross section[15]. Like optic fibres the difference in effective index between the "core" (unoxidised) and "cladding" (fully oxidised) regions is small. The $HE_{11}$ mode in our undoped cavity has quality factors between 6000 and 8000 in pillars where we estimate the size of the aperture to be 1-2 µm. Similar mode structures were also observed in a microcavity containing a tapered aperture beginning at an electric field antinode (data not shown here) despite the fact that the radial variation in effective index takes a different form. However, the similar optical confinement exhibited by the two different structures confirmed that the spacings and positions of the modes are a function of aperture diameter, the absolute change in refractive index and not the precise form by which the effective index varies.

Figure 2(b) shows data taken from another 12 µm mesa. Here the cavity mode was observed around 933.0 nm at 4 K together with a QD emission line (identified as a neutral exciton) at 932.6 nm. By increasing the temperature, it was possible to tune the QD through the cavity mode. Figure 2(c) shows the measured lifetime of the exciton state as a function of temperature. At 27 K, when the dot and cavity are on resonance, the lifetime falls to a minimum value of ~ 450 ps. By comparison to the lifetime of this dot when detuned from the cavity mode, this corresponds to a Purcell factor of 2.4. It is well known that the proportion of emission which is directed into the mode by the Purcell Effect is given by $\beta = (F-f)/F$ where F is the Purcell Factor and $f$ is a function of the decay rate into leaky modes and non-radiative paths. In the case of this cavity $f \sim 1$ implying $\beta \sim 60\%$. Given that the lateral extent of the mode is ~ 1-2 µm in this cavity, several times the mode wavelength, diffraction should be small and it should be possible to achieve a high collection efficiency. We were, however, not able to determine the collection efficiency experimentally as it was not possible to observe saturation in the intensity of the exciton line due to emission from other states at high excitation powers.

We now discuss a second structure, Wafer 2, which consisted of a similar cavity but with only 4 (12) mirror periods. The upper (lower) mirror was Si- (Be-) doped to form a p-i-n diode, similar to our previous work[13]. An array of devices were etched and oxidised to produce apertures of ~1 µm diameter. Ohmic contacts were then made to the doped layers using standard techniques.

Figure 3(a) shows optical spectra from one device in which a single QD was observed in the cavity mode. The upper line plots the measured reflectivity profile recorded at 4 K. A broad cavity mode was observed, centred on 915 nm with $Q \sim 100$. This is less that the calculated value of 150 due to the finite range of angles over which emission was collected. Current could be injected into the device above a voltage of 2.0V and electroluminescence (EL) was observed from a number of QD states. The lower line in figure 3(a) plots the EL spectrum from the device at 3.0 V on the same wavelength scale. The most intense emission line, at 916.4 nm, was close to the center of the cavity mode and had a linear intensity dependence with injected current, as shown in figure 3(b). Measurements showed a small wavelength splitting between H- and V- polarised components of the state. Such a splitting, which arises from a physical, in-plane asymmetry in the dot, is typically observed in the neutral emission states from these QDs. These observations indicate that the feature is likely to originate from a neutral exciton (X) state.

The emission statistics were studied using a Hanbury-Brown and Twiss experiment consisting of a beamsplitter, two avalanche photodiodes and a time-correlated counting card. The recorded histogram for direct current operation at 100 nA (figure 3(c)) shows sub-Poissonian photon emission with $g^{(2)}(0)$ of 0.21, confirming that the majority of the emission originates from a single quantum state. The experimental data have been modelled using a rate-equation model[16] which assumes transitions are purely radiative. We also include the effect of the finite response time of the experiment (550ps), together with the experimentally determined background emission levels. This fit gives a good agreement with the experimental data (solid line, figure 3(c)).

In conclusion, we have demonstrated that a tapered oxide aperture can be used to achieve optical and electrical confinement in a semiconductor microcavity. In an optically pumped structure, emission from an InAs quantum dot was coupled to the cavity mode and a Purcell factor of 2.4 was observed. In a p-i-n diode from a doped structure, EL was observed from a single QD with antibunching. We conclude that this device design is promising for generating high-efficiency single-photon emission.

The authors would like to acknowledge support from the European Commission under the Integrated Project SECOQC, the Integrated Project Qubit Applications (QAP) funded by the IST directorate as Contract Number 015848 and Framework Package 6 Network of Excellence SANDiE. One of the authors (D.J.P.E) would also like to thank EPSRC and Toshiba for funding.

**Figure Captions**

Figure 1: Top: schematic of the electrical device. Not to scale. Electrical contact is made through the upper and lower mirrors. The linear taper of the oxide layer extends over 1 µm. Bottom: calculated radial distribution of the effective refractive index in the device, on the same scale as the schematic.

Figure 2: (a) photoluminescence spectrum showing the optical modes of an undoped structure under high excitation density. (b) a plot of the wavelength of the $HE_{11}$ mode and a quantum dot exciton state as a function of temperature in one of these pillars. As the detuning between the exciton and mode is reduced the lifetime falls, as expected from the Purcell effect, shown in (c).

Figure 3: (a) reflectivity from the doped cavity structure and (lower trace) electroluminescence excited at 3.0V. The intensity of the emission line at 916.4 nm displays a linear dependence of the current injected to the device (b) suggesting it is excitonic. Under direct current injection, photons emitted from this single state are antibunched with $g^{(2)}(0) = 0.21$ (c).

**Figure 1**

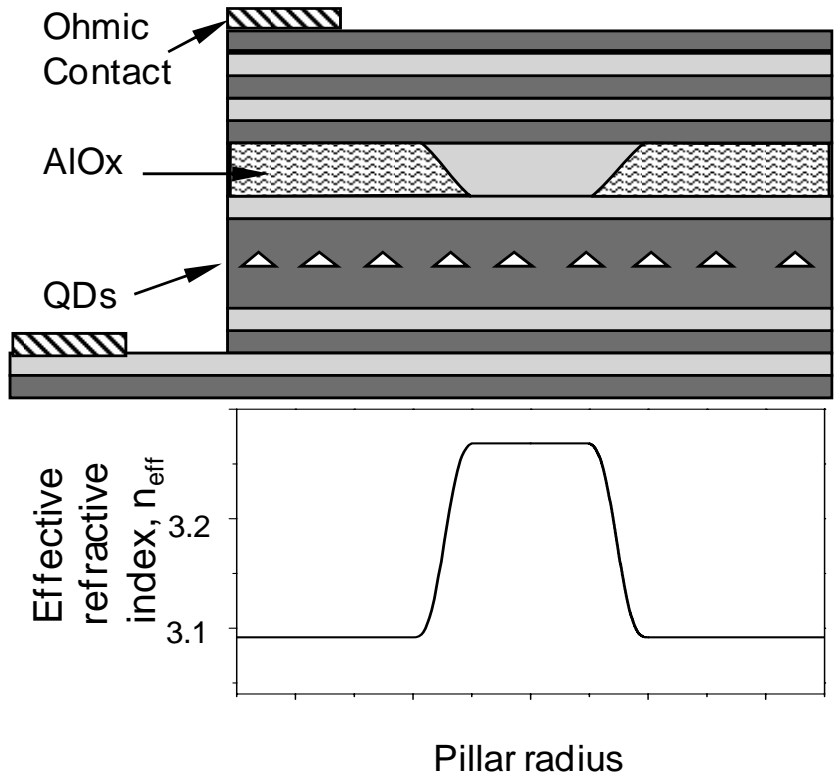

**Figure 2**

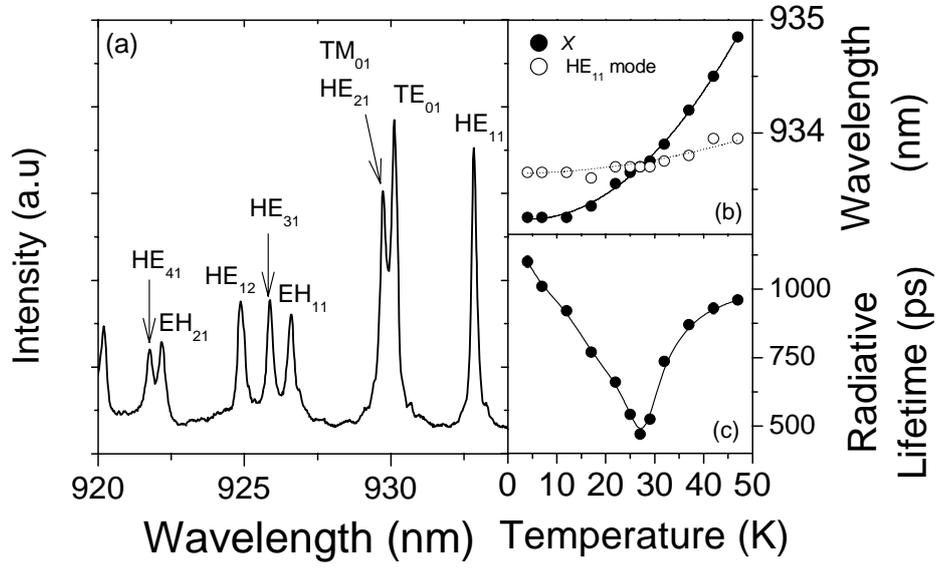

**Figure 3**

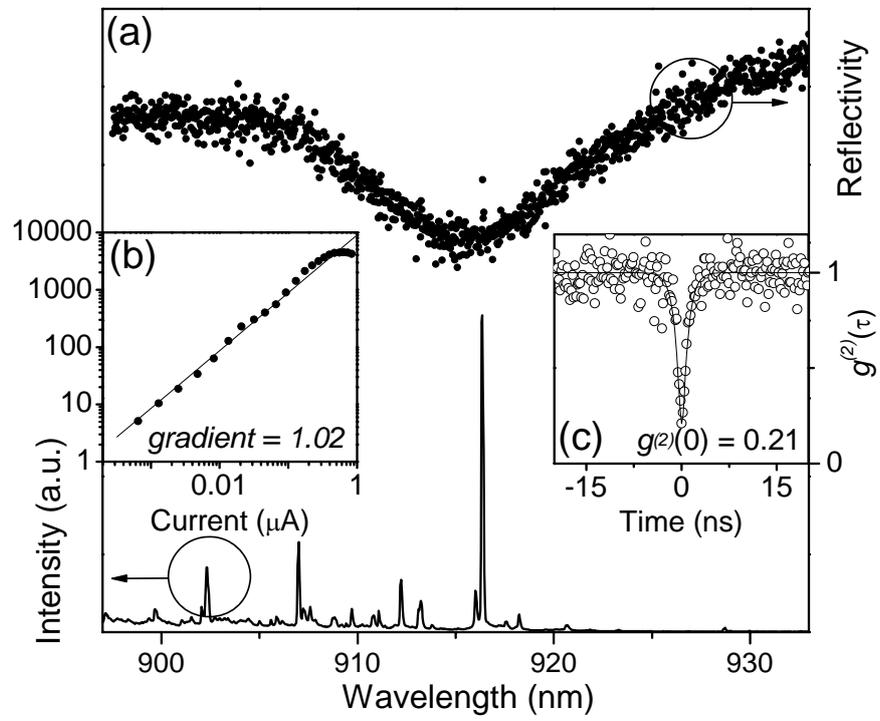